\documentclass[prl,showpacs,superscriptaddress,amsmath,amssymb,twocolumn,aps]{revtex4}
\usepackage{graphicx}
\usepackage{dcolumn}
\usepackage{bm}

\begin{document}

\title{Superconductivity in a single layer alkali-doped FeSe: a weakly coupled two-leg ladder system}

\author{Wei Li}
\affiliation{State Key Laboratory of Low-Dimensional Quantum Physics, Department of Physics, Tsinghua University, Beijing 100084, China}
\author{Hao Ding}
\affiliation{State Key Laboratory of Low-Dimensional Quantum Physics, Department of Physics, Tsinghua University, Beijing 100084, China}
\author{Pengfei Zhang}
\affiliation{State Key Laboratory of Low-Dimensional Quantum Physics, Department of Physics, Tsinghua University, Beijing 100084, China}
\author{Peng Deng}
\affiliation{State Key Laboratory of Low-Dimensional Quantum Physics, Department of Physics, Tsinghua University, Beijing 100084, China}
\author{Kai Chang}
\affiliation{State Key Laboratory of Low-Dimensional Quantum Physics, Department of Physics, Tsinghua University, Beijing 100084, China}
\author{Ke He}
\affiliation{Institute of Physics, Chinese Academy of Sciences, Beijing 100190, China}
\author{Shuaihua Ji}
\affiliation{State Key Laboratory of Low-Dimensional Quantum Physics, Department of Physics, Tsinghua University, Beijing 100084, China}
\author{Lili Wang}
\affiliation{Institute of Physics, Chinese Academy of Sciences, Beijing 100190, China}
\author{Xucun Ma}
\affiliation{Institute of Physics, Chinese Academy of Sciences, Beijing 100190, China}
\author{Jian Wu}
\affiliation{State Key Laboratory of Low-Dimensional Quantum Physics, Department of Physics, Tsinghua University, Beijing 100084, China}
\author{Jiang-Ping Hu}
\affiliation{Institute of Physics, Chinese Academy of Sciences, Beijing 100190, China}
\author{Qi-Kun Xue}
\email{qkxue@mail.tsinghua.edu.cn} 
\affiliation{State Key Laboratory of Low-Dimensional Quantum Physics, Department of Physics, Tsinghua University, Beijing 100084, China}
\author{Xi Chen}
\email{xc@mail.tsinghua.edu.cn}
\affiliation{State Key Laboratory of Low-Dimensional Quantum Physics, Department of Physics, Tsinghua University, Beijing 100084, China}

\date{\today}

\begin{abstract}

We prepare single layer potassium-doped iron selenide (110) film by molecular beam expitaxy. 
Such a single layer film can be viewed as a two-dimensional system composed of weakly coupled two-leg iron ladders. 
Scanning tunneling spectroscopy reveals that superconductivity is developed in this two-leg ladder system. 
The superconducting gap is similar to that of the multi-layer films. However, the Fermi surface topology given by first-principles calculation
is remarkably different from that of the bulk materials. 
Our results suggest that superconducting pairing is very short-ranged
or takes place rather locally in iron-chalcogenides. The superconductivity is most likely driven by 
electron-electron correlation effect and is insensitive to the change of Fermi surfaces.

\end{abstract}


\maketitle

In spite of intensive investigation, there is still no consensus on whether a strong-coupling or a weak-coupling model 
is more appropriate for describing high temperature superconductivity. 
For the recently discovered iron-based superconductors \cite{hosono08,xhchen08,wang08a,xlchen10}, 
such an issue is still an open debate as well \cite{wang08b,mazin08,aoki08,lee09,
bernevig09,bernevig11,chubukov08,tesanovic09,hu08a,si08,hu08b,xiang08,hu12a,scalapino10,hu12b}. 
In the strong-coupling picture, the essential physics of high Tc superconductivity is expected to be understood in terms of localized models. 
Alternatively, a weak-coupling model begins by considering the energy bands and Fermi surface. 
One approach to addressing this problem is to discover and study high Tc superconductors with reduced dimensionality. 
For example, quasi-one-dimensional systems, such as ladder or chain materials, 
are simple model systems for theories of superconductivity based on magnetic pairing mechanisms \cite{rice96a,rice94,rice96b,scalapino02}. 
Experimentally, the only spin-ladder system found to be superconducting so far is Sr$_{14-x}$Ca${_x}$Cu$_{24}$O$_{41}$ (x = 11.5-15.5) 
\cite{kinoshita96,tsunetsugu97}. However, the origin of superconductivity in this ladder material is still controversial 
because it becomes superconducting only by applying high pressure, which may change one-dimensional physics. 
Here we report on the superconductivity in a single layer K$_x$Fe$_{2-y}$Se$_2$(110) film. 
The single layer K$_x$Fe$_{2-y}$Se$_2$(110) is a two-dimensional system composed of weakly coupled two-leg iron ladders. 
The finding indicates that superconducting pairing is very short-ranged 
and most likely driven by electron-electron correlation effect in iron-chalcogenides.

The single layer K$_x$Fe$_{2-y}$Se$_2$ along the [110] direction was grown by molecular beam epitaxy (MBE). 
The (001) plane is the natural cleavage plane of the layered material K$_x$Fe$_{2-y}$Se$_2$ 
and the (110) plane shows the cross-section. 
The (001) oriented single layer film can be viewed as a two-dimensional system 
composed of weakly coupled two-leg iron ladders [Fig. 1(a)]. 
Using scanning tunneling microscopy (STM) and spectroscopy (STS), 
we demonstrate that a superconducting gap is developed and the gap size is reduced by only 10\% 
compared with that of a multi-layer film \cite{chen12a}. The prominent advantage of the single layer K$_x$Fe$_{2-y}$Se$_2$ 
over other ladder systems is that this system is obtained by reducing the dimensionality 
(especially the in-plane dimensionality) of an existing bulk superconductor. 
The two-leg Fe ladder can be considered as the building block of both bulk and single layer K$_x$Fe$_{2-y}$Se$_2$, 
and provides an opportunity to study the dimensionality effect and therefore casts new lights on resolving the controversy 
between strong-coupling and weak-coupling models. By studying the superconductivity in this ladder system, 
we show that superconducting pairing in iron-based superconductors stems 
from local short-ranged microscopic energetics rather than the Fermi surface properties.

\begin{figure*}[t]
        \includegraphics[width=5.5in]{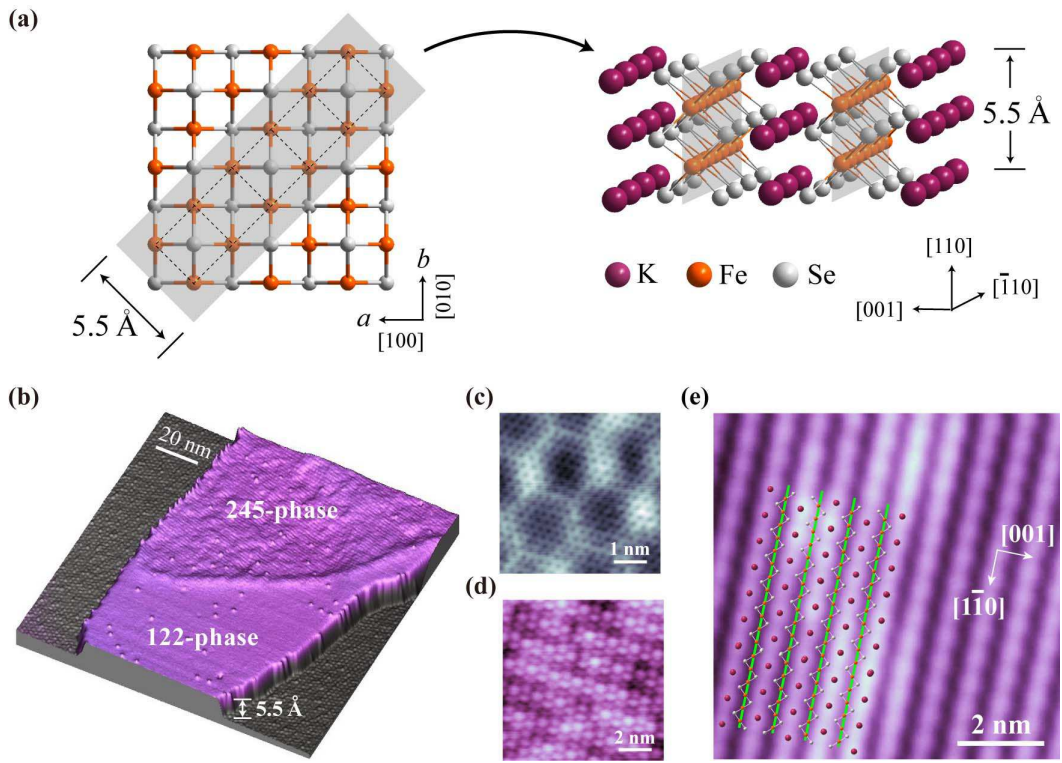}
        \caption{\label{fig1}  Single layer K$_x$Fe$_{2-y}$Se$_2$  (110). 
        (a) Two-leg ladder system composed of weakly coupled iron atom chains. The same convention for Miller indices is used throughout.  
        In the single layer two-leg ladder system (schematic on right), each FeSe(001) layer contributes one ladder 
        (the shaded area in the schematic on left, dashed lines marked the Fe ladder), which is along the [$\bar{1}$10] direction. 
        A layer of K atoms is sandwiched between two ladders.  
        (b) STM topographic image (150 nm $\times$ 150 nm, 3.9 V, 0.02 nA) of a single layer K$_x$Fe$_{2-y}$Se$_2$(110) island. 
        Two distinct regions are marked by 122-phase and 245-phase. The height of island is 5.5 {\AA} and is in agreement 
        with the schematic in (a).
        The substrate is bi-layer graphene characterized by 6$\times$6 hexagonal superstructure (c). (d) STM image of 245-phase 
        (1.35 V, 0.03 nA). The periodic stripe pattern is attributed to the $\sqrt{5}\times\sqrt{5}$ Fe vacancy order 
        seen along the [110] direction.
        (e) STM image of 122-phase (8 nm $\times$ 8 nm, 10 mV, 0.02 nA). The locations of iron ladders are marked by dashed lines. 
         The locations of K and Se atoms are marked by red and white dots, respectively. }
\end{figure*}

The experiments were performed on a Unisoku ultra-high vacuum STM system at base temperature of 0.4 K 
by means of a single-shot $^3$He cryostat. The system has a MBE chamber for thin-film growth. 
A magnetic field up to 11 T can be applied perpendicular to the sample surface. 
A polycrystalline PtIr STM tip was used in the experiments.

The single layer (110) films of K$_x$Fe$_{2-y}$Se$_2$(110) [Fig. 1(b)] were obtained on graphitized 6H-SiC(0001) 
under well-controlled growth conditions. The substrate [dark region in Fig. 1(b)] is bi-layer graphene, 
characterized by the $6 \times 6$ hexagonal superstructure \cite{veuillen07} [Fig. 1(c)] and a typical gap-like feature 
in $dI/dV$ spectrum \cite{crommie08} (see Supplementary Material \cite{supplimentary}). 
Similar to our previous studies \cite{chen12a,chen12b}, 
two distinct regions known as insulating 245-phase with $\sqrt{5}\times\sqrt{5}$ Fe vacancy order [Fig. 1(d)] 
and conducting 122-phase [Fig. 1(e)] are clearly revealed on the island, 
indicating that phase separation occurs at the single layer level as well. 
The two phases always coexist side by side. In a single layer, 
the composition of K$_x$Fe$_{2-y}$Se$_2$(110) in the 122-phase actually 
becomes K$_3$Fe$_4$Se$_6$ instead of KFe$_2$Se$_2$. But nominally we still use ``122'' to label this conducting phase.

We focus on the 122-phase, which is superconducting in the case of multi-layer films. 
Along the [110] direction, a single layer K$_3$Fe$_4$Se$_6$(110) contains 
one complete unit cell of 122-phase and its thickness is 5.5 {\AA} [see Fig. 1(a) and (b)]. 
The (110) plane is terminated with K and Se atoms. 
The K atoms form atomic rows and are visible at positive bias in the STM image [Fig. 1(e)]. 
Each of the weakly coupled two-leg iron ladders is located in the middle between two adjacent K atom rows. 
This unique configuration provides an ideal platform to explore superconductivity in ladder systems.

\begin{figure}[h]
        \includegraphics[width=3.25in]{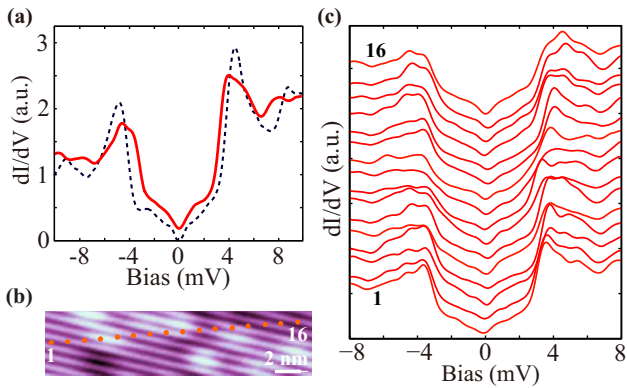}
        \caption{\label{fig2}   Superconductivity in single layer 122-phase. 
        (a) The superconducting gap (solid curve) of single layer 122-phase. Setpoint: 14 mV and 0.1 nA. Temperature: 0.4 K.
        For comparison, the spectrum for multi-layer 122-phase (dashed curve) is shown.  
        (b) A 5 nm $\times$ 25 nm topography image (14 mV, 0.1 nA) of the superconducting region. 
        Inhomogeneity in electronic structure is revealed. (c) A series of $dI/dV$ spectra measured at 16 points indicated in (b).  
        Setpoint: 14 mV and 0.1 nA. The spectra are offset for clarity.}
\end{figure}

STS probes the local density of states of quasiparticles. In Fig. 2(a), 
the STS measurements on single layer 122-phase at 0.4 K show clear evidence of 
superconductivity in this two-leg ladder system. The superconducting gap is centered at the Fermi level 
and has two coherence peaks. Similar to the spectrum acquired on multi-layer sample 
[dashed curve in Fig. 2(a) and also Ref. \cite{chen12a}], the STS reveals a double-gap structure. 
The coherence peaks for the single layer 122-phase are slightly broadened 
and the superconducting gap is reduced from 4.0 meV for the multi-layer film to 3.6 meV. 
The spectra always exhibit a finite zero bias conductance most likely because of the 
quasiparticle scattering from the substrate. The thickness of the film is only 5.5 {\AA} and 
much shorter than the coherence length of 122-phase in the a-b plane. 
Thus the influence of substrate is not negligible.

While structural defects are rarely found in the single layer film, 
inhomogeneity in electronic structure commonly exists as shown in STM image [Fig. 2(b)]. 
A series of $dI/dV$ spectra taken along the dotted line in Fig. 2(b) also exhibit small variation 
of superconducting gap at different locations. Such inhomogeneity does not present 
in multi-layer films \cite{chen12a} and can be attributed to the substrate.

\begin{figure}[h]
        \includegraphics[width=3in]{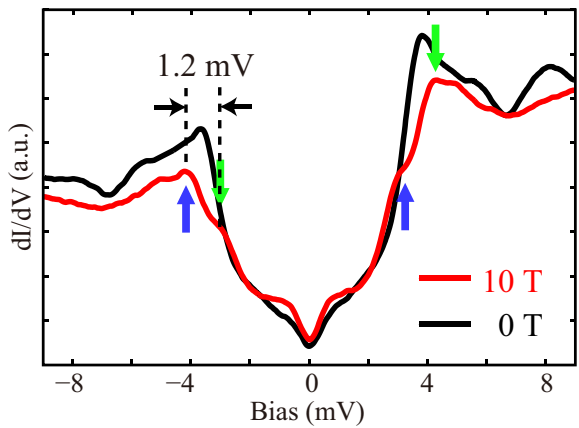}
        \caption{\label{fig3}  Magnetic field splitting of the quasiparticle states. 
        $dI/dV$ spectra show density of states for single layer 122-phase at 0 T and 10 T, respectively. Setpoint: 14 mV, 0.1 nA. 
        The coherence peaks are split into spin-up and spin-down components under magnetic field. 
        }
\end{figure}

Besides the substrate, the superconducting gap is susceptible to external magnetic field as well [Fig. 3]. 
When a magnetic field is applied parallel to a very thin film or to a type II superconductor, 
the effect of field on spins dominates the one on orbits \cite{fulde11}. 
At the Pauli limit, where Zeeman energy is comparable to the superconducting gap, 
the Copper pairs are broken. Below the Pauli limit, the spectrum simply shifts by a Zeeman term in energy. 
One coherence peak splits into the up and down spin states and the splitting is given by $2\mu_BB$, 
where $\mu_B$ is the Bohr magneton. 
This effect has been observed experimentally in thin superconducting aluminum films \cite{fulde70,meservey71,tedrow86,meservey91,adams96}. 
In the single layer 122-phase, the splitting at 10 T is 1.2 meV [Fig. 3], close to the theoretical value 1.16 meV. 
The same behavior is obtained at different locations and on multi-layer film (see Supplementary Material \cite{supplimentary}).

To discern the key electronic structure that is responsible for superconductivity, 
we performed first-principles calculation [Fig. 4] within the density function theory 
using the projected augmented-wave method \cite{blochl94} as implemented in the VASP code \cite{kresse96}. 
The Perdew-Burke-Ernzerhof exchange correlation potential \cite{perdew96} was used. 
The calculation confirms that the single layer 122-phase is a quasi-one-dimensional metal 
with two bands [Fig. 4(a)] crossing the Fermi surface. 
The ground-state is antiferromagnetic (see Supplementary Material \cite{supplimentary}) 
with a large magnetic moment of 3.284 $\mu_\text{B}$  per Fe atom. 
The Fermi surface [Fig. 4(b)] is very different from those of bulk iron based superconductors. 
On the boundary of Brillouin zone, there are one electron pocket at N point and one hole pocket at M points. 
They are highly elongated along the direction perpendicular to the ladders. 
In particular, there is no hole pocket at $\Gamma$ point. The band structure calculation indicates 
that the Fermi surface topology in the momentum space is not crucial to the formation of Cooper pairs.

\begin{figure}[h]
        \includegraphics[width=2.75in]{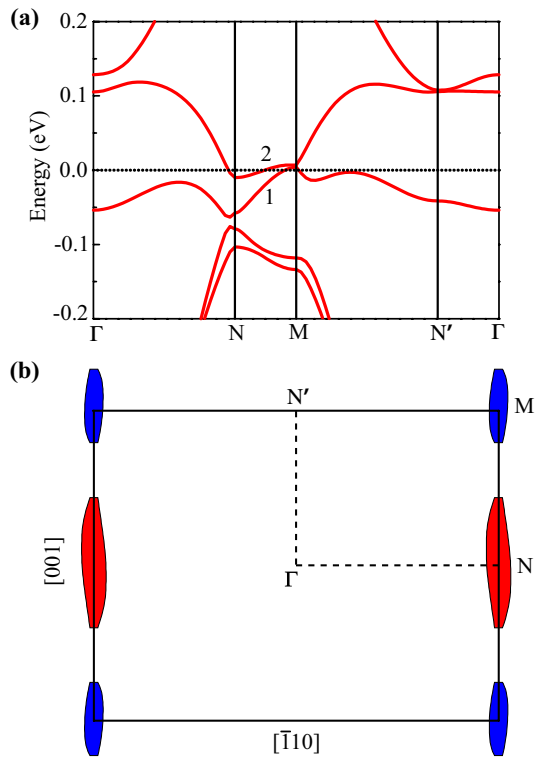}
        \caption{\label{fig4}  First-principles calculation of single layer K$_3$Fe$_4$Se$_6$.
        A 500 eV cutoff in the plane-wave expansion ensures the calculations converge to $10^{-5}$ eV. 
        For each magnetic configuration, all atomic positions and the lattice constants were optimized 
        until the largest force on each atom is 0.01 eV/\AA. 
        A 12$\times$12$\times$1 Monkhorst-Pack k-grid Brillouin zone are sampled throughout all calculation 
        while the Gaussian smearing technique is used in the case of metallic states. 
        To model the thin film, a supercell of slab is used with periodic boundary conditions 
        in all the three dimensions with a 10 {\AA}  thick vacuum layer between two slabs in order to eliminate the inter-slab interactions. 
        (a) The band structure. There are two bands crossing the Fermi surface. (b) Fermi surface sheets. 
        The electron and hole pockets are located at N and M points respectively. The Fermi energy sets to zero.
                }
\end{figure}

Our results on single layer two-leg ladder system have important implications. 
First of all, the development of superconductivity in such a low dimensional system
indicates that superconducting pairing is very short-ranged or takes place rather locally 
in iron-chalcogenides. The superconductivity is most likely driven by electron-electron correlation effect 
and is insensitive to the Fermi surface topology. The results strongly support 
earlier theoretical calculations performed on iron-pnictide superconductor in a two-leg 
ladder model \cite{scalapino10,scalapino09}, which suggests that the underlying pairing mechanism for iron-pnictide 
superconductors is similar to that for the cuprates. Although those calculations were done for iron-pnictides, 
there is a good reason to expect their validity for iron-chalcogenides without significant modification. 
For example, recently it has been shown that the underlining kinematics between iron-pnictides 
and iron-chalcogenides has little difference and the similarity between iron-based superconductors 
and cuprates can be extended to a fully two dimensional model \cite{hu12b}. 
Finally, the combination of the theoretical investigations and our experimental results reveals 
that the pairing in iron-based superconductors is dominated by the interaction between 
two next nearest neighbor iron sites. A two-leg ladder system is 
exactly the minimum one-dimensional model to capture the essential physics of iron-based superconductors.

In summary, we observe superconductivity in a single layer alkali-doped iron selenide,
which is a nature-made weakly-coupled two-leg ladder system. 
Our results indicate that the material is a strongly correlated electron system with very short-ranged local superconducting pairing.   
  
The work is supported by NSFC and the National
Basic Research Program of China. The STM topographic
images were processed using WSxM (www.nanotec.es).

\end{document}